\documentclass[prl,twocolumn]{revtex4}

\usepackage{graphicx}
\usepackage{bm}
\usepackage{ifthen}

\begin{document}

\title{Experimental interference of independent photons}

\author{Rainer Kaltenbaek}
\affiliation{Institute for Experimental Physics, University of Vienna,
Boltzmanngasse 5, A-1090 Vienna, Austria}
\author{Bibiane Blauensteiner}
\affiliation{Institute for Experimental Physics, University of Vienna,
Boltzmanngasse 5, A-1090 Vienna, Austria}
\author{Marek \.Zukowski}
\affiliation{Institute for Experimental Physics, University of Vienna,
Boltzmanngasse 5, A-1090 Vienna, Austria}
\affiliation{University of Gdansk, Institute of Theoretical Physics and
Astrophysics, Wita Stwosza 57, PL-08-952 Gdansk, Poland}
\author{Markus Aspelmeyer}
\affiliation{Institute for Experimental Physics, University of Vienna,
Boltzmanngasse 5, A-1090 Vienna, Austria}
\affiliation{Institute for Quantum Optics and Quantum Information
(IQOQI), Austrian Academy of Sciences, Boltzmanngasse 3, A-1090 Vienna,
Austria}
\author{Anton Zeilinger}
\affiliation{Institute for Experimental Physics, University of Vienna,
Boltzmanngasse 5, A-1090 Vienna, Austria}
\affiliation{Institute for Quantum Optics and Quantum Information
(IQOQI), Austrian Academy of Sciences, Boltzmanngasse 3, A-1090 Vienna,
Austria}

\pagestyle{empty}

\begin{abstract}
Interference of photons emerging from independent sources is essential
for modern quantum information processing schemes, above all quantum
repeaters and linear-optics quantum computers. We report an observation
of non-classical interference of two single photons
originating from two independent, separated sources, which were actively 
synchronized with an r.m.s.~timing jitter of $260$~fs. The resulting (two-photon) interference visibility was $83\pm 4$~\%. 
\end{abstract}

\maketitle

Is it possible to observe fully destructive interference of
photons if they all originate from separate, independent sources? Yes,
according to quantum theory \cite{Dicke1964a,Mandel1983a,Paul1986a}. 
The perfect interference of photons 
emerging from independent sources cannot be
understood by the classical concept of the superposition of
electromagnetic fields but only by the interference of probability
amplitudes of multi-particle detection events. As
stressed by Mandel ``this prediction has no classical analogue, and its
confirmation would represent an interesting test of the quantum theory
of the electromagnetic field'' \cite{Mandel1983a}. 

Mastering the techniques involving independent sour\-ces of single photons
and entangled pairs of photons has become vital for implementations of
quantum networks and quantum computing schemes
\cite{Bouwmeester2000a,Nielsen2000a}. For these devices to work it is
often tacitly assumed that stable interference between systems from
independent sources is feasible. The generic example is that of quantum
repeaters \cite{Briegel1998a}, which by definition involve entanglement
swapping and distillation between spatially separated, independent nodes
requiring independent sources. Teleportation of states of
particles emitted by sources completely detached from the sources of the
entangled pairs of the quantum channel could become feasible. Other
applications are linear optics quantum computing schemes of the KLM-type
\cite{Knill2000a}, in which ancilla qubits need to become entangled to
other, independent optical qubits during the process of the computation.

To demonstrate that two independently emitted photons do interfere, it
is important to assure that there exists no possibility whatsoever for the 
coherence properties of the light emitted by either source to be influenced 
by the other. Therefore, the operation of one source must not in any
way rely on the working of the other source. Such a configuration
addresses exactly the needs for practical quantum communication and
computation schemes. In the case of long-distance quantum communication
any common optical elements shared by the sources and thus any
dependence would impede the working of the scheme over large distances
due to dispersion or losses. Our experiment fulfills these requirements
for independent quantum sources. At the same time it serves as a
prototype solution for a variety of quantum information processing
devices. 

First, consider two independent classical sources. Any correlation between
intensities at two detectors placed in the joint far-field of the sources
is a manifestation of standard
interference of classical waves and shows at most 50~\% 
visibility \cite{Paul1986a}.
This is only observable if the detector integration times are below the
coherence times of the two fields. A well-known example is the stellar
interferometry method introduced by Hanbury-Brown and Twiss
\cite{HanburyBrown1956a}. 

The situation becomes fundamentally different for quantum states of
light, e.g. in the case of two separate spontaneously decaying atoms. While one
photon can be detected practically anywhere, there are points for which
detection of the second photon is then strictly forbidden. The resulting
correlation pattern has $100 \%$ visibility, completely unexplainable by
interference of classical waves. This is due to destructive interference of
two indistinguishable processes: (a) the photon registered in the first
detector came from source 1 and the photon registered in the second
detector from source 2, and (b) the photon registered in the first
detector came from source 2 and the photon registered in the second
detector from source 1.

Quantum interference of two fully independent photons has thus far never
been observed. Since the 1960s, however, interference of light from 
independent sources has been  addressed in many experiments. 
In \cite{Javan1962a} two independent He-Ne lasers
were used to observe the beating of their superposed outputs. 
Later \cite{Magyar1963a}, transient spatial interference fringes between 
beams from independent ruby lasers were reported. In both cases the
interference was classically explainable. Partly motivated by the often
overinterpreted quotation from Dirac that each photon interferes only
with itself \cite{Dirac1958a}, follow-up experiments 
\cite{Pfleegor1967a,Radloff1971a} investigated the
question whether one can observe interference of two photons if
each one was generated by a different source. This was done by simply 
attenuating the laser beams. However, attenuation does not affect the 
statistical nature of laser light. The only quantum aspect was that the 
detection involved clicks due to photon registrations. Consequently, the
observed effects could ``not readily be  described in terms of one photon 
from one source interfering with one from the other'' \cite{Pfleegor1967a}.

All following experiments involving the interference between single photons
employed the well-known Hong-Ou-Mandel (HOM) interference effect, 
which utilizes the bosonic nature of photons: two indistinguishable photons 
that enter a 50:50 beam splitter via different input ports will always be 
detected in one output port. Such two-photon interference was first 
reported  \cite{Hong1987a} for photon pairs 
emerging from a spontaneous parametric down-conversion (SPDC) source. 

The first interference of separately generated 
photons was observed by Rarity et al.~\cite{Rarity1996a} (see 
also~\cite{Kuzmich2000a}). They measured
Hong-Ou-Mandel-type (HOM) interference \cite{Hong1987a} of an SPDC photon 
and an attenuated part of the very same laser beam pumping the SPDC process. 
Further related experiments, provided gradual progress with respect to the 
independence of the utilized sources. A first step was the interference of two 
triggered single photons created via SPDC by the same pump pulse passing 
twice through the very same SPDC crystal \cite{Bouwmeester1997a}. Further 
contributions used photons generated by two mutually coherent time-separated 
pulses from the same mode-locked laser in one SPDC crystal \cite{Keller1998a} 
and, later, generated in one quantum dot \cite{Santori2002a}.  Another 
step was to create interfering photons in two separate SPDC 
crystals pumped by the same laser \cite{deRiedmatten2003a}. The most 
recent experiment along that line used pulses from two intersecting 
laser cavities sharing the same Kerr medium \cite{Yang2005a}.

However, as has been pointed out in one of those prior works, 
``truly independent sources require the use of independent but synchronized 
fs laser[s]'' \cite{deRiedmatten2003a}. Our experiment employs this technique
and realizes a scheme involving two independent quantum sources which can 
in principle be separated by large distances.

The photons emitted from a quantum source are typically generated by the 
interaction of an (optical) pump field with a nonlinear medium. The medium 
and the pump field are integral constituents of the source. In our experiment, 
each of the two sources consists of an SPDC crystal pumped optically 
by a pulsed fs laser. 

To be able to observe interference we have to make sure that the two
photons registered behind the beam splitter cannot be distinguished in any way. 
We use SPDC to generate pairs of correlated photons. The detection event
of one of the photons (trigger) of each pair is used to operationally define 
the presence of the other one on its way to the beam splitter 
(in this way we assure that the observed interference is due 
to two photons only, each from a different source). In such a case without 
frequency filtering, the initial sharp time correlation of photons of 
an SPDC pair poses a problem: the times of registration of the 
trigger photons provide temporal distinguishability of the
photon registrations behind the beam splitter. Short pump pulses and
spectral filters narrower than the bandwidth of these pulses in the paths of 
the photons give the desired indistinguishability~\cite{Zukowski1995a}. 

\begin{figure}[ht]
  \begin{center}
  \includegraphics[width=0.9\linewidth]{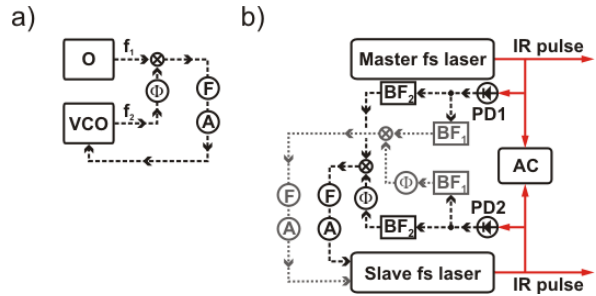}
  \caption{\textbf{(a)} A phase-locked loop (PLL) synchronizes a voltage 
controlled oscillator (VCO) relative to another oscillator (O). The 
frequency of the VCO is adjusted by the feedback signal of a phase detector
($\bigotimes$), which is fed through a low-pass filter (F) and an 
amplifier (A) (see e.g,~\cite{Horowitz1989a}).
\textbf{(b)} The pump lasers are time-synchronized by a Coherent 
Synchrolock$^{TM}$ using two PLLs. One operates at the repetition 
frequency of the lasers ($76$~MHz) for a coarse time-synchronization. 
Then, this PLL is switched off and the second PLL operating at the 
lasers' $9^{th}$ harmonic ($684$~MHz) takes over. Both PLLs adjust the 
``slave'' laser's repetition frequency via cavity mirrors driven by piezo 
actuators. The PLLs are fed by fast photo diodes ($PD_1$ and $PD_2$) 
filtered by bandwidth filters ($BF_1$ and $BF_2$) to get the fundamental 
and $9^{th}$ harmonic signals. The performance of the synchronization is
observed via an autocorrelator (AC).
  }
  \label{SYNC}
  \end{center}
\end{figure}

Additional timing information is contained in the time difference
between the independent pulses pumping the two SPDC crystals. In principle, 
one could compensate this again by filtering. For pulses
without any time correlation this would, however, require extremely
narrow filters and eventually result in prohibitively low count rates.
Synchronizing the pulses of the two independent pumps increases the 
probability of joint emission events (see fig. \ref{SYNC}b) and
hence the count rates. \textit{The fact that one needs to actively
synchronize the sources is a direct unavoidable consequence of their 
independence.} 
The active synchronization method we use involves only electronic 
communication ($10$~kHz bandwidth) about the relative pulse timing between 
the independently running femtosecond lasers (see fig. \ref{SYNC}b). 
No optical elements whatsoever are shared by the pumps. 

\begin{figure}[ht]
  \begin{center}
  \includegraphics[width=0.65\linewidth]{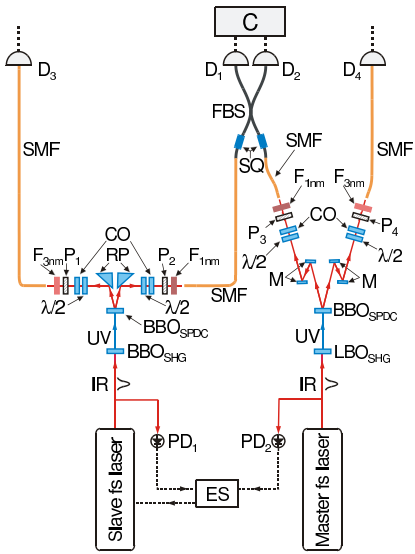}
  \caption{
Pulsed IR laser beams, which were electronically synchronized (ES, see 
fig.~\ref{SYNC}), were frequency-doubled (one in a Lithium-Triborate~(LBO), 
the other in a $\beta$-Barium Borate~(BBO) crystal). The resulting UV beams 
pumped type-II BBO-crystals for SPDC. Reflecting prisms (RP) and mirrors (M) 
guided the SPDC photons through half-wave plates and BBO 
crystals (CO) to compensate various walk-off effects. All photons were 
coupled into single mode fibers (SMF) to guarantee optimal spatial mode
overlap. Polarizers $P_1$-$P_4$, narrow bandwidth filters $F_1$-$F_4$ and 
fiber squeezers (SQ) ensured the indistinguishability of the photons at 
the single-mode fiber beamsplitter. Coincidences C between the detectors 
$D_1$ and $D_2$ could be triggered on detection events in both $D_3$ and 
$D_4$.
  }
  \label{SETUP}
  \end{center}
\end{figure}

Our two SPDC crystals were pumped by UV pulses with
centre wavelengths of $394.25\pm 0.20$~nm and
$394.25\pm 0.20$~nm and r.m.s.~bandwidths of $0.7\pm 0.1$~nm and
$0.9\pm 0.1$~nm. These beams were produced via 
frequency doubling of IR pulses from two independent Ti:Sa femtosecond lasers
(master and slave, see fig. \ref{SETUP}). One of these mode-locked lasers
was driven by an Ar-Ion gas laser, the other by a solid-state Nd:YAG laser.
They produced pulses at approx. $76$~MHz repetition rate with centre
wavelengths of $788.5\pm 0.4$~nm and $788.5\pm 0.4$~nm, r.m.s.~bandwidths 
of $2.9\pm 0.1$~nm and $3.2\pm 0.1$~nm and r.m.s.~pulse 
widths of $49.3\pm 0.3$~fs and $46.8\pm 0.3$~fs. 
The laser pulses were synchronized via electronic feedback loops up to a 
relative timing jitter of $260\pm 30$~fs using the commercially available 
Synchrolock$^{TM}$ system from Coherent Inc. (see fig.~\ref{SYNC}b).

To observe the interference of two independent photons, we varied the
time delay between the two lasers in $300$~fs steps with an accuracy
better than $100$~fs. The measurement time for each data point was $900$~s.
Long-time drifts of the relative delay between the lasers were compensated
by measuring in blocks of $60$~s and by automatical readjustment of the 
delay between these measurement blocks.
This was done by tuning the intensity of the light detected by one of the
fast photo diodes used for synchronization, which introduces a small change
of delay between the lasers, which was monitored via an 
autocorrelator (AC).

\begin{figure}[ht]
  \begin{center}
  \includegraphics[width=0.9\linewidth]{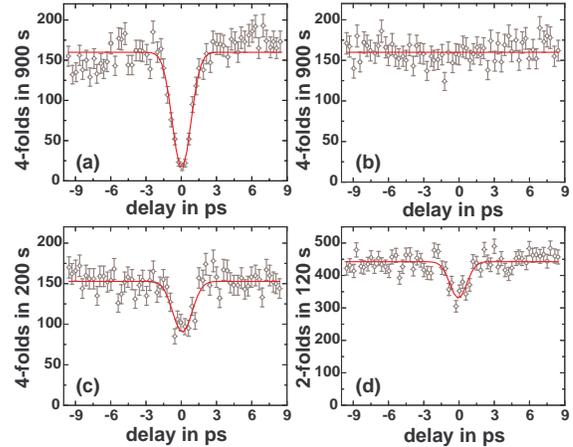}
  \caption{
Experimental two-photon interference from independent sources. 
\textbf{(a)} HOM-type interference of indistinguishable photons from 
actively synchronized independent sources. The observed visibility was
$83\pm 4$~\% and the dip width was $0.79\pm 0.04$~ps. 
\textbf{(b)} Input photons distinguishable by their polarization. No 
interference occurs.
\textbf{(c)} Unpolarized input photons show limited interference due to 
partial distinguishability. The observed visibility was ($26\pm 3$)~\%. 
\textbf{(d)} Classical interference from a thermal source, showing a
dip visibility of $15\pm 2$~\%.
  }
  \label{RESULTS}
  \end{center}
\end{figure}

The interference, in the form of a Hong-Ou-Mandel dip, is shown in Figure 
\ref{RESULTS}a. The visibility of $83\pm 4$~\% is well beyond the 
classical limit of $50 \%$ \cite{Paul1986a}. Both the observed visibility 
and the r.m.s.~dip width of $0.79\pm 0.03$~ps agree well with the theoretically 
expected values of $84\pm 3$~\% and $0.86\pm 0.07$~ps, given the relative pulse 
timing jitter and filter bandwidths (see Appendix). Our result therefore 
clearly agrees with the quantum predictions.

To additionally demonstrate the role played by
distinguishability in this effect we prepared different input states
under otherwise equivalent experimental conditions. First, we used
perfectly distinguishable orthogonally polarized input states, which as 
expected show no interference (Fig. \ref{RESULTS}b). Next, unpolarized 
input photons (Fig. \ref{RESULTS}c) were used 
which are a mixture of orthogonally polarized photons and hence are partially 
distinguishable. They still have a probability of $\frac{1}{2}$ to coincide in 
their polarization, which results in an expected visibility of ideally 
$33$~\% or, taking into account the relative timing jitter, $29.6\pm 0.8$~\%. 
We observed $26\pm 3$~\%. Finally, we demonstrated the interference for 
photon sources endowed with thermal statistics. Without monitoring the trigger 
detection events, the emission statistics in each input mode a and b of the 
beam splitter is equivalent to light emitted by a thermal source. For two such 
beams of equal average intensity one would expect $20$~\% visibility 
\cite{Paul1986a} in the ideal case or $18.0\pm 0.5$~\% when bearing in mind 
the relative timing jitter. Experimentally we achieved a lower visibility of 
$14.5\pm 2.0$~\% due to differences of the SPDC-pair rates in the two sources
(approx. a factor of $2$). 
Note, that for specially prepared classical light sources the visibility can 
even reach the very maximum of $50$~\% \cite{Paul1986a}.

Our experiment demonstrates the feasibility of interference of two
single photons originating from independent, spatially separated sources, which 
were actively time-synchronized. The visibility of the effect 
is above the threshold for further use in quantum communication processes
like quantum teleportation or entanglement swapping. This result is a step 
towards the realization of quantum repeaters, quantum networks and certain 
optical quantum computing schemes. Due to the separation of the utilized
sources the presented scheme opens the door for future long distance 
applications involving multi-photon interference. Moreover, the use of such 
independent sources might also provide conceptual advantages for experiments 
on the foundations of quantum physics \cite{Yurke1992a}.

We are grateful to \v C.~Brukner, A.~Dominguez, Th.~Jennewein, A.~Poppe, 
K.~Resch, P.~Walther and G.~Weihs for theoretical discussion and 
experimental advice. We acknowledge support from the Austrian Science 
Fund FWF (SFB F15), the DTO-funded U.S.~Army Research Office Contract 
No.~W911NF-05-0397 and the European Commission under projects RamboQ, 
SECOQC and QAP. MZ is supported by an FNP Professorial Subsidy and MNiI 
grant Nr.~1P03~04927. The collaboration is supported by the 
Austrian-Polish Collaboration Program \"oAD/MNiI. 

\appendix*

\section{APPENDIX}

To obtain the theoretical expectations for the HOM-dip via standard quantum 
electrodynamics. We assume both lasers 
to have an r.m.s.~bandwidth of $\sigma_p$, both interfering photons to be 
filtered to an r.m.s.~bandwidth $\sigma_S$ and both trigger photons
to $\sigma_T$. The timing jitter between the two generated 
SPDC pairs is given by $\sigma_J=350\pm 30$~fs, resulting from the jitter 
of the laser synchronization ($260\pm 30$~fs gaussian jitter) and the 
group-velocity mismatch between UV and IR photons in the SHG and SPDC crystals. 
The central wavelengths of the lasers and the filters are assumed to be equal. 

With these assumptions the visibility of the HOM dip is given by
\begin{equation}
\frac{ \sigma_p }{ 2 \sqrt{ \frac{(\sigma^2_S+\sigma^2_p+2 \sigma^2_p\sigma^2_J \sigma^2_S) (\sigma^2_p + \sigma^2_T)}{\sigma^2_p + \sigma^2_S + \sigma^2_T}} - \sigma_p},
\end{equation}
which reduces to the formula given in \cite{Zukowski1995a} for $\sigma_J=0$
and $\sigma_T \to \infty$.

By the same method the dip width is found to be
\begin{equation}
\frac{\sqrt{\sigma^2_p + \sigma^2_S 
(1+2 \sigma^2_J \sigma^2_p)}}{\sqrt{2} \sigma_S \sigma_p}.
\end{equation}
A detailed derivation, also for more general cases, is given 
elsewhere~\cite{Kaltenbaek2006b}. 

\bibliography{rk}

\end{document}